\documentclass[aps,prl,twocolumn,superscriptaddress, nobibnotes,showpacs]{revtex4-1}
\usepackage{gensymb}
\usepackage{graphicx}
\usepackage{xcolor}
\usepackage{ulem}
\usepackage{bm,hyperref}

\begin{document}

\title{\textbf{Fermi-Surface Instabilities in the Heavy-Fermion Superconductor UTe$_{2}$}}

\author{Q. Niu}
\affiliation{Univ. Grenoble Alpes, CEA, IRIG, PHELIQS, F-38000 Grenoble, France}
\author{G. Knebel}
\affiliation{Univ. Grenoble Alpes, CEA, IRIG, PHELIQS, F-38000 Grenoble, France}
\author{D. Braithwaite }
\affiliation{Univ. Grenoble Alpes, CEA, IRIG, PHELIQS, F-38000 Grenoble, France}
\author{D. Aoki }
\affiliation{Univ. Grenoble Alpes, CEA, IRIG, PHELIQS, F-38000 Grenoble, France}
\affiliation{Institute for Materials Research, Tohoku University, Oarai, Ibaraki, 311-1313, Japan}
\author{G. Lapertot}
\affiliation{Univ. Grenoble Alpes, CEA, IRIG, PHELIQS, F-38000 Grenoble, France}
\author{G. Seyfarth}
\affiliation{Laboratoire National des Champs Magn\'etiques Intenses (LNCMI), CNRS, Univ. Grenoble Alpes, 38042 Grenoble, France}
\author{J-P. Brison}
\affiliation{Univ. Grenoble Alpes, CEA, IRIG, PHELIQS, F-38000 Grenoble, France}
\author{J. Flouquet}
\affiliation{Univ. Grenoble Alpes, CEA, IRIG, PHELIQS, F-38000 Grenoble, France}
\author{A. Pourret}
\email[E-mail me at: ]{alexandre.pourret@cea.fr}
\affiliation{Univ. Grenoble Alpes, CEA, IRIG, PHELIQS, F-38000 Grenoble, France}
\date{\today } 

\begin{abstract}

We present different transport measurements up to fields of 29~T in the recently discovered heavy-fermion superconductor UTe$_{2}$ with magnetic field $H$ applied along the easy magnetization a-axis of the body-centered orthorhombic structure. The thermoelectric power varies linearly with temperature above the superconducting transition, $T_{SC}= 1.5$ K, indicating that superconductivity develops in a Fermi liquid regime. As a function of field the thermolelectric power shows successive anomalies which are attributed to field-induced Fermi surface instabilities. These Fermi-surface instabilities appear at critical values of the magnetic polarization. Remarkably, the lowest magnetic field instability for $H\parallel a$ occurs for the same critical value of the
magnetization (0.4 $\mu_B$) than the first order metamagnetic transition at 35~T for field applied along the $b$-axis. The estimated number of charge carriers at low temperature reveals a metallic ground state distinct from LDA calculations indicating that strong electronic correlations are a major issue in this compound.
\end{abstract}


\maketitle

Unconventional superconductivity (SC) in heavy-fermion systems is the consequence of  the delicate interplay between competing magnetic and non-magnetic ground states. Recent studies on uranium based ferromagnetic superconductors have pointed out the interplay between magnetic fluctuations and a Fermi surface (FS) reconstruction on crossing the quantum phase transition at the ferromagnetic to paramagnetic instability \cite{Aoki2019a}. The emergent picture is that the change of the amplitude of the ferromagnetic correlations and even the switch of the direction of the magnetic fluctuations are directly associated with a FS instability. A reinforcement of SC (RSC) has been observed in the three uranium based ferromagnetic superconductors URhGe \cite{Aoki2001}, UCoGe \cite{Huy2007a} and UGe$_2$ \cite{Saxena2000}. In particular, in URhGe and UCoGe RSC appears when a magnetic field is applied along the hard-magnetization $b$-axis and it is linked to the increase of magnetic and electronic fluctuations due to the collapse of the Curie temperature in a field transverse to the easy magnetization axis. FS instabilities induced by a magnetic field, such as Lifshitz transitions \cite{Lifshitz1960}, have been observed in these heavy fermion materials \cite{Yelland2011,Gourgout2016, Aoki2014,Bastien2016} underlining the importance of the Zeeman splitting of the flat bands crossing the Fermi level. The respective role of such FSs instabilities on the mechanism of RSC and the sole magnetic fluctuations is still an open question\cite{Yelland2011,Sherkunov2018}.

The recently discovered heavy-fermion superconductor UTe$_2$ \cite{Ran2018, Aoki2019} with $T_{sc}$=1.6~K, is one of the rare examples of heavy-fermion materials with a superconducting transition temperature above 1~K. In contrast to ferromagnetic UCoGe and URhGe, UTe$_2$ is a paramagnetic material but nevertheless it exhibits RSC \cite{Ran2018, Knebel2019, Ran2019b} up to unrivalled magnetic field strengths among this class of materials. 
 UTe$_2$ has an orthorhombic crystal structure. The easy magnetization axis is the $a$-axis, and the $c$-axis is the hard axis above 20~K \cite{Ran2018}. For $H \parallel b$, there is a maximum of the magnetic susceptibility at $T_{\chi~max} \approx 35$~K, so that at $T=2$~K the susceptibility is lowest for the $b$-axis. This maximum of the susceptibility is related to the first order metamagnetic transition at $H_m =35$~T, which has been observed in recent high field magnetization \cite{Miyake2019, Ran2019b} and resistivity experiments \cite{Knafo2019, Ran2019b} for field applied along the hard magnetization $b$-axis. The superconducting upper critical field $H_{c2}$ is very anisotropic: $H^a_{c2}=6$~T, $H^c_{c2}=12$~T,  and $H^b_{c2}=H_m=35$~T for a magnetic field applied along the $a$, $c$, and $b$ axis, respectively. The values of $H_{c2}$ for all directions highly exceed the Pauli limit. Spectacularly, when a magnetic field is applied along the $b$-axis, the $H_{c2}(T)$ exhibits S-shape curves with a strong reinforcement on approaching $H_m$ \cite{Knebel2019}.
 The unconventional shape of $H_{c2}$ can be modelled assuming a field-induced enhancement of the pairing strength in the strong coupling limit \cite{Aoki2019, Knebel2019}.
 
The connection of a FS instability with the field-enhancement of SC has been most conclusively documented for URhGe \cite{Yelland2011, Gourgout2016}. In UCoGe FS instabilities have been reported for the easy magnetization $c$-axis which have been identified as Lifshitz transitions linked to a critical value of the magnetization \cite{Bastien2016}. However, RSC appears along the $b$-axis in this system possibly also linked to another FS instability \cite{Malone2012}.
%
The FS of UTe$_2$ has not been determined up to now.  No quantum oscillations have been observed and  recent photoemission spectroscopy experiments were not able to resolve the electronic band structure close to the Fermi level \cite{Fujimori2019}. The first LDA band structure calculation obtained a Kondo semiconducting ground state with very flat bands near the Fermi energy \cite{Aoki2019}. These results do not correspond to the real metallic electronic states observed at low temperature in UTe$_2$. By shifting the $5f$ level upward by 0.2 Ry in the LDA calculations, small FSs which occupy only 5\% of the Brillouin zone appear \cite{Harima2019}, suggesting that UTe$_2$ would be a semi-metallic system with heavy electronic states.
In this Letter, we focus on the transport properties of  UTe$_2$ under magnetic field applied along the easy magnetization $a$-axis. In this direction, the magnetization above $T_{sc}$ increases nonlinearly with the field, showing a tiny change of slope around 6.5~T, and starts to saturate above $H\approx 21$~T  (reaching $1.05~\mu_B$ at 40~T) \cite{Miyake2019}. We observe in the different transport properties a series of anomalies as a function of magnetic field applied along the $a$ axis which can be assigned to Lifshitz transitions. The first Lifshitz transition, $H_1$, occurs at same value of magnetization, 0.4 $\mu_B$, at which the metamagnetism occurs for $H \parallel b$. Our experiments show that UTe$_2$ is a good metal with about one charge carrier per U atom. 

Single crystals of UTe$_2$ were grown by chemical vapor transport with iodine as transport agent. The orientation of the crystals has been verified by Laue diffraction.  In order to study the field dependence of the FS, we performed thermoelectric power ($S$) resistivity ($\rho$) and thermal conductivity ($\kappa$) measurements on three different samples (labeled S1, S2, S3) with residual resistivity ratios ($\frac{\rho(300~K)}{\rho(1.6~K)}$) of 30, 16 and 38. They have been prepared for experiments with heat or electric current along the $a$-axis. $S$, $\kappa$ and $\rho$ were measured with field along the $a$-axis. The Hall effect was determined by applying the field along the $b$ and $c$-axis, respectively. The temperature and field dependence of the different transport properties have been measured on sample S1 in CEA Grenoble using a home-made dilution refrigerator with a base temperature of 100 mK and a superconducting magnet with field up to 16 T and on sample S2 in a standard PPMS above 4K with field up to 9T. Furthermore, we performed measurements at LNCMI Grenoble using a $^3$He cryostat up to 29~T on sample S1. $S$ and $\kappa$ have been measured using a standard "one heater-two thermometers" setup compatible with dc resistivity measurements. The Hall effect was measured on S2 ($H \parallel b$) and S3 ($H \parallel c$) in a PPMS.

\begin{figure}
\includegraphics[width=0.5\textwidth]{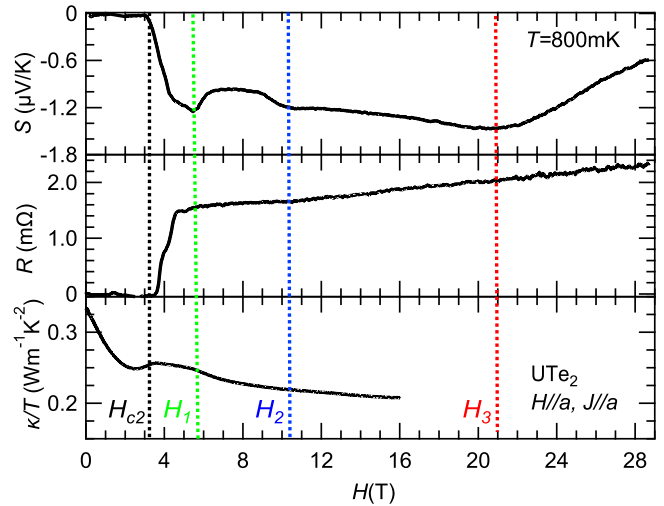}
\caption{(color online) Comparison of the magnetic field dependence of  $S$, $\rho$ and $\kappa/T$ at 800~mK in UTe$_2$ (S1). The different critical fields are represented by dashed vertical lines.}
\label{Fig1}
\end{figure}

Figure \ref{Fig1} compares the magnetic field dependence   of the different transport coefficients $S$, $\rho$ and $\kappa/T$ at 800~mK for $H \parallel a$. Different successive anomalies occur. $H_{c2}$ defined by $S=0$ or $\rho=0$ corresponds to a sharp kink in $\kappa/T$, and the onset of  SC in $S$ and $\rho$ coincides perfectly. Above $H_{c2}$, $S(H)$ is negative and exhibits a sharp peak at $H_1 = 5.6$~T. At that field $\kappa/T (H)$ has a distinct change of slope, only a tiny change of the magnetic susceptibility has been observed around 6.6~T \cite{Miyake2019}. For even higher fields additional anomalies appear, most pronounced in $S (H)$ at $H_2=10.5$~T and $H_3 = 21$~T. The anomaly at $H_2$, corresponding to a tiny increase of $\rho$, has been already identified previously \cite{Knafo2019}. $H_3$ is only visible in $S$ but appears to be linked to the entrance into the saturating regime of magnetization under field \cite{Miyake2019}.

\begin{figure}
\includegraphics[width=0.5\textwidth]{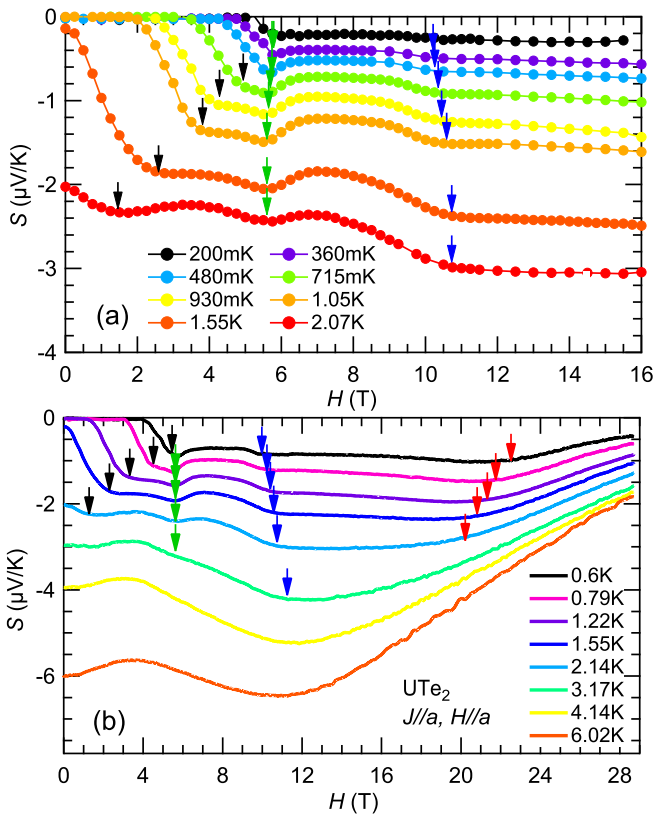}
\caption{(color online) Field dependence of $S$ in UTe$_2$ (S1) for $H \parallel a$, (a) below 2~K up to 16~T measured using a superconducting magnet by averaging every point,  and (b) below 6~K up to 29~T by sweeping continuously the magnetic field at LNCMI on sample S1. The arrows indicate the onset of the superconducting state and the position of the electronic instabilities .}
\label{Fig2}
\end{figure}

Figure 2 focuses on the thermoelectric power for $H\parallel a$: panel (a) shows the field dependence of  $S$ up to 16~T and panel (b) up to 29~T for different temperatures. Three anomalies can be observed above the superconducting critical field and followed as a function of temperature: the marked minimum at $H_1= 5.6$~T is independent of temperature and merges with $H_{c2}$ at very low temperature. The anomaly at $H_2$ and the broad one at $H_3$ depends more on temperature. They can be followed up to 3~K, but get less pronounced. Above 3~K we observe a broad maximum around 4~T and a minimum around 12~T.  

\begin{figure}
\includegraphics[width=0.5\textwidth]{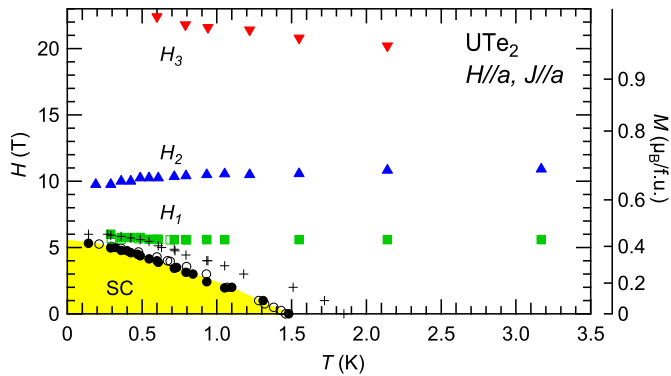}
\caption{(color online) Magnetic-field temperature phase diagram of  UTe$_2$ (S1) for $H \parallel a$. The upper superconducting critical field $H_{c2}$ determined by $S = 0$, the onset of the superconducting transition (crosses),  and the different critical fields, $H_1$, $H_2$ and $H_3$, observed in $S$ (full symbol), $\kappa$ (open symbol) and $\rho$ (crosses) are represented. These critical fields correspond to specific values of magnetization (right scale) measured at $T_{sc}$ at 1.4~K from Ref. \onlinecite{Miyake2019}.}
\label{Fig3}
\end{figure}

$H_{c2}(T)$ and the different anomalies $H_1$, $H_2$ and $H_3$ are displayed in the magnetic-field temperature phase diagram in Fig. \ref{Fig3}. Using magnetization from Ref.~\onlinecite{Miyake2019}, the corresponding magnetization scale, measured just above $T_{sc}$, is represented on the right axis. The key phenomenon is that FS changes are induced by crossing some critical values of magnetic polarization. 
We see that $H_1$ occurs when the magnetization reaches $M \approx 0.4~\mu_B$/f.u. Remarkably, for $H \parallel b$ the magnetization is of the same order just below the magnetization jump at $H_m=35$~T \cite{Miyake2019} which may indicate a critical magnetization value. At $H_3$, $M(H)$ starts to saturate.
 
\begin{figure}
\includegraphics[width=0.5\textwidth]{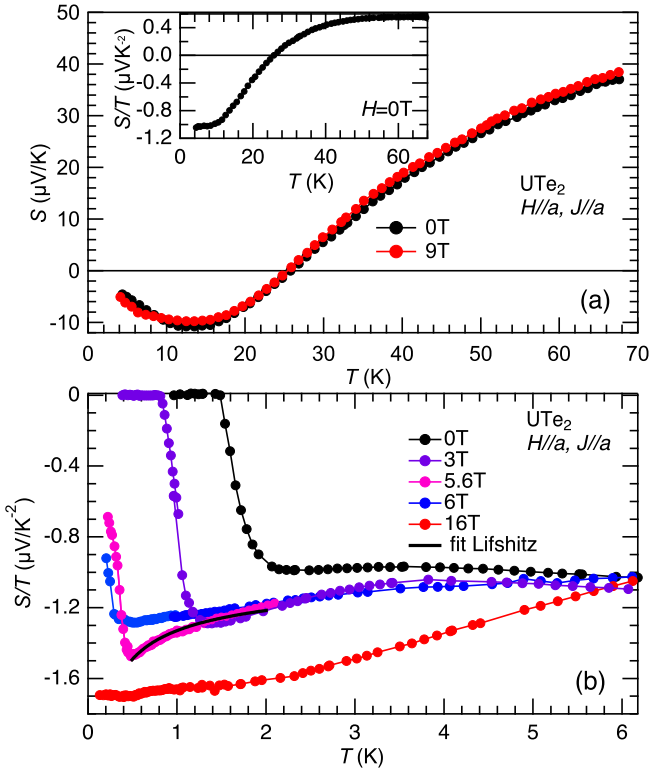}
\caption{(a) Temperature dependence of $S$ at $H$=0 and 9~T above 4~K (S2). The inset shows the temperature dependence of $S/T$ for $H=0$~T. (b) Temperature dependence of $S/T$ for different magnetic field below 6~K (S1). $S/T$ at 5.6~T is fitted using a temperature dependence expected for a Lifshitz transition, $T^{-\frac{1}{2}}$.}.
\label{Fig4}
\end{figure}

The temperature dependence of  $S$ is represented in Fig.~\ref{Fig4}(a). For $H=0$, $S$ is positive at high temperature, changes sign at 25~K, where the hardest magnetization axis changes from $c$ to $b$ axis and the longitudinal magnetic fluctuations start to develop along the $a$-axis \cite{Tokunaga2019}. Decreasing temperature, $S$ shows a broad minimum at around 12~K and goes linearly to zero below 10~K, as shown by the constant value of $S/T=-1 \mu$V/K$^2$ (see inset of Fig.~\ref{Fig4}). A constant $S/T$ strongly supports a Fermi liquid regime at low temperature in UTe$_2$ as already observed in resistivity measurements \cite{Aoki2019, Ran2018}. In contrast,  it is not sufficient to explain the singular temperature dependence of the magnetization measured for $H \parallel a$ \cite{Ran2018, Miyake2019}, notably the strong increase at low temperature. Thus, as in many 5f compounds, a decoupling appears between the magnetic response dominated by the local U moments and the quasiparticles at the Fermi level.
The field dependence of $S(T)$ is very weak, at least up to 9~T in this temperature range. $S(T)/T$ below 6~K is represented in Fig.~\ref{Fig4}(b). In zero-field, $S/T$ extrapolated to 0~K gives -$1\mu$V/K$^2$. For $H=H_1=5.6~$T, $S/T$ shows a downward curvature exhibiting the $T^{-\frac{1}{2}}$ dependence in $\frac{S/T_{(H=5.6)}}{S/T_{(H=0)}}$ expected for a Lifshitz transition \cite{Varlamov1989}. The fit, $a+(T^*/T)^{-\frac{1}{2}}$ which gives a characteristic energy $T^*=150$~mK, is represented by a black line. Lifshitz transitions \cite{Lifshitz1960} correspond to the appearance or disappearance of  small FS pockets inside the Brillouin zone. It is now agreed that the main consequence on transport properties is not the change in the density of states but the change of the scattering rate \cite{Blanter1994, Varlamov1989}. Such a topological change of the FS will act as a trap for electrons in the scattering process from main land FS (large $k$) to small pockets (small $k$) through impurities. So, the peak in $S$ is most probably a field-induced Lifshitz transition occurring at $H_1$. For $H=6$~T, just above $H_1$, $S/T$ is again constant at low temperature. 
In many heavy fermion systems field-induced Lifshitz transitions have now been identified \cite{Daou2006, Pourret2013, Pourret2013a, Pfau2013a,  Aoki2016, Bastien2016, Pfau2017}, but not in every case they are related to a metamagnetic transition. The case of UTe$_2$ can be compared to the series of FS reconstructions observed in the ferromagnetic superconductor UCoGe when applying a field along the easy magnetization $c$-axis, where up to five anomalies ($H_1$-$H_5$) have been detected. The Lifshitz character of the transitions in UCoGe has been confirmed by the observation of changes in the quantum oscillations frequencies \cite{Bastien2016}. In UCoGe, magnetization measurements for $H \parallel a$ do not show any detectable metamagnetic transition, especially not at the most marked anomaly at $H_4$, where the Hall effect changes sign from positive to negative. However, this anomaly in UCoGe occurs also for a magnetization of $M \approx 0.4~\mu_B/$f.u.. For $H \parallel b$,which is the hardest axis in UCoGe, metamagnetism occurs at the same value of magnetization at $H=45$~T.

In many correlated "good" metals, the absolute value of the dimensionless ratio $q=\frac{SN_{Av}e}{\gamma T}$ ($\gamma$ is the Sommerfeld coefficient of the specific heat, $e$ the elementary charge, and $N_{Av}$ the Avogadro number) is of the order of unity \cite{Behnia2004a}. It has been argued that, in the zero-temperature limit, for scattering both in the Born and unitary limits, $S/T$ becomes inversely proportional to the renormalized Fermi energy, and this leads to the observed correlation \cite{Miyake2005}. Let us recall that when the carrier density is much lower than one itinerant electron per formula unit, a proportionally larger $|q|$ is expected. $S/T \approx -1$~$\mu$V/K$^2$ and $\gamma \approx 0.12$~J/(K$^2$mol) \cite{Aoki2019} yield $q\approx -0.8$, giving $n_U=\frac{1}{|q|}\approx1.2$ carriers (electrons) per formula unit. This simple argument indicates that, as regards the value of $S/T$ extrapolated at 0~K, the zero field ground state of UTe$_2$ is an heavy fermion metal with a significant number of charge carriers. 

\begin{figure}[h!]
\includegraphics[width=0.5\textwidth]{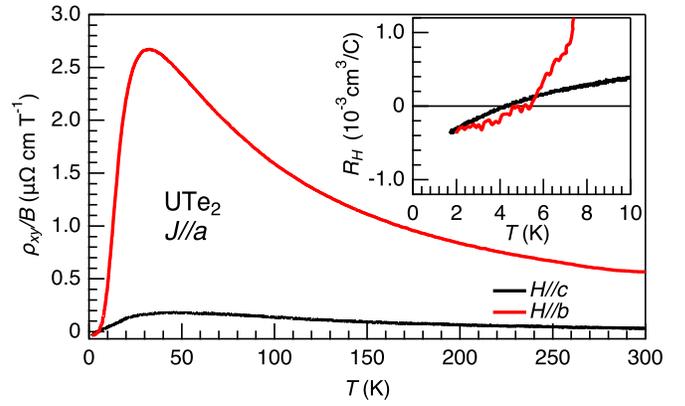}
\caption{$\rho_{xy}/H$ as a function of temperature measured at 9~T for $H\parallel b$ (S2) and $H\parallel c$ (S3). The inset shows a zoom of $R_H$ in the low temperature region.}
\label{Fig5}
\end{figure}

To further check the number of carriers in UTe$_2$, we have measured the Hall resistivity $\rho_{xy}$. In Fig.~\ref{Fig5}, $\rho_{xy}/H$ as a function of temperature measured at 9~T is represented for two magnetic field orientations, $H\parallel b$ and $H\parallel c$. For both field directions, $\rho_{xy}/H$ shows a pronounced maximum around $T_{\chi~max}\approx 35$~K and changes sign below 5~K, see inset of Fig.~\ref{Fig5}. For $H \parallel c$, the Hall effect is much smaller,  
but the low temperature value at 2~K is similar, $\rho_{xy} \approx-36$~ n$\Omega$ cmT$^{-1}$. The extrapolated value of the Hall resistance at $T_{sc}=1.5$~K, $R_H=-3.8\times10^{-4}$ cm$^3$/C (corresponding to $\rho_{xy}=39$~n$\Omega$cmT$^{-1}$) where the influence of the skew scattering at $T_{\chi~max}$ can be neglected, gives a density of electrons $n= 1.6\times 10^{22}$cm$^{-3}$. Using a volume per formula unit of  $V_U=88.9$~\AA$^3$, this yields a number of carriers (electrons) per formula unit of  $n_U=1.4$. Another way to estimate the carrier number is to use thermodynamic data. The  Sommerfeld coefficient $\gamma$ is proportional to $\frac{k^2_F}{v_F}$ and $k_F \propto n^{1/3}$. The Fermi velocity $v_F$ can be estimated from the slope of $H_{c2}$ at $T_{sc}$ and we get $v_F \propto (\frac{dH_{c2}(T_{sc})}{dT})^{-\frac{1}{2}}$, and thus the charge carrier number $n=(\gamma v_F)^{\frac{3}{2}}$. The slope of $H_{c2}$ at $T_{sc}$ and in consequence $v_F$ is sensitive to the field orientation, $v_F=11000$,  5500 and 9500 m/s for $H \parallel a, b, c$ respectively \cite{Aoki2019}. Taking the average value of $v_F$ and $\gamma \approx 0.12$~J/(K$^2$mol) yields $n_U=0.51$ carriers per uranium atom. This value is slightly lower than the values estimated from the transport but still classifies UTe$_2$ as a good metal. 

However, UTe$_2$ is a compensated metal, and we expect equal number of holes and electrons. Band structure calculations predict FSs from the hole and electron bands \cite{Harima2019, Fujimori2019}, and the valence of U-atoms is very sensitive to the Coulomb interaction $U$ \cite{Ishizuka2019}. As the initial slope of $H_{c2}$ is determined from the heaviest charge carriers and the transport properties are equally sensitive to the light charge carriers, we expect that the FS is created from heavy holes and light electrons. Recently, photoemission spectroscopy experiments have been reported \cite{Fujimori2019}. Part of these experiments agree with the calculated band-structure, however, the details of the band structure close to the Fermi level could not be resolved. The observation of an incoherent peak in the photoemission spectra has not been predicted. This indicates the importance of strong correlation effects which require to be treated beyond the LDA approach to resolve the paradigm of a large number of charge carriers.

In conclusion, the number of charge carriers in UTe$_2$ extracted from $S$, $\rho_{xy}$ and from the link between $H_{c2}$ and $\gamma$ establishes a metallic and highly correlated ground state distinct from a Kondo insulator ($\approx$ 1e$^-/U$). A series of  FS instabilities have been identified in different transport properties for field along the $a$-axis. The temperature dependence of the anomaly in $S$ at  $H_1$ confirms clearly the topological nature of  this transition. The values of the critical polarization at which $H_1$ occurs for $H\parallel a$ and at which $H_m$ occurs for $H \parallel b$ are remarkably similar indicating the strong interplay between magnetic polarization and electronic instabilities in this system.

\begin{acknowledgments}
The authors thank K. Izawa, S. Hosoi, H. Harima, J. Ishizuka, Y. Yanase for stimulating discussions. This work has been supported by the University Grenoble Alpes and KAKENHI (JP15H0082, JP15H0084, JP15K21732, JP19H00646, JP16H04006, JP15H05745). We acknowledge support of the LNCMI-CNRS, member the European Magnetic Field Laboratory (EMFL).
\end{acknowledgments}

\bibliographystyle{apsrev4-1}	

%
\end{document}